\documentclass[12pt]{article}
\usepackage{amsfonts,amstext,epsfig}

\let\a=\alpha \let\b=\beta \let\g=\gamma \let\d=\delta 
   \let\k=\kappa
     
\let\s=\sigma \let\t=\tau

\let\pa=\partial
\def\be{\begin{equation}}
\def\ee{\end{equation}}
\def\ba{\begin{array}}
\def\ea{\end{array}}

\def\dalemb#1#2{{\vbox{\hrule height .#2pt
        \hbox{\vrule width.#2pt height#1pt \kern#1pt
                \vrule width.#2pt}
        \hrule height.#2pt}}}

\newcommand{\bea}{\begin{eqnarray}}
\newcommand{\eea}{\end{eqnarray}}
\newcommand{\tr}{{\rm tr} }
\newcommand{\Tr}{{\rm Tr} }

\def\bZ{{{\Bbb Z}}}

\newcommand{\preprint}[1]{\begin{table}[t]  %%
             \begin{flushright}               %%
             {#1}                             %%
             \end{flushright}                 %%
             \end{table}}                     %%
\renewcommand{\title}[1]{\vbox{\center\LARGE{#1}}\vspace{5mm}}
\renewcommand{\author}[1]{\vbox{\center#1}\vspace{5mm}}
\newcommand{\address}[1]{\vbox{\center\em#1}}
\newcommand{\email}[1]{\vbox{\center\tt#1}\vspace{5mm}}

\begin{document}

\begin{titlepage}
\preprint{hep-th/0603190 \\
DAMTP-2006-21}

\title{Multiply wound Polyakov loops at strong coupling}

\author{Sean A. Hartnoll$^1$ and S. Prem Kumar$^2$}

\address{$^1$DAMTP, Centre for Mathematical Sciences,
Cambridge University\\
Wilberforce Road, Cambridge CB3 OWA, UK \\
\vspace{0.2cm}
$^2$Department of Physics, University of Wales Swansea \\
Swansea, SA2 8PP, UK}

\email{s.a.hartnoll@damtp.cam.ac.uk, s.p.kumar@swansea.ac.uk}

\abstract{
We study the expectation value of a Polyakov-Maldacena loop that
wraps the thermal circle $k$ times in strongly coupled
${\mathcal{N}}=4$ super Yang-Mills theory. This is achieved by
considering probe D3 and D5 brane embeddings in the dual black
hole geometry. In contrast to multiply wound spatial Wilson loops,
nontrivial dependence on $k$ is captured through D5 branes. We find
$N^{-2/3}$ corrections, reminiscent of the scaling behaviour near a
Gross-Witten transition.}

\end{titlepage}

\section{Motivation and Introduction}

\subsection{Phase structure of ${\mathcal{N}}=4$ SYM theory}

The phase structure of ${\mathcal{N}}=4$ super Yang-Mills (SYM) theory at
large $N$ on $S^3 \times S^1$ is of interest from both
gravitational and field theoretic perspectives.
The AdS/CFT correspondence \cite{ads/cft}
places both the weak and strong coupling regimes of the theory within
computational reach, suggesting the possibility of fully mapping
out the phase structure as a function of temperature and coupling.

At strong coupling it was observed
\cite{Witten:1998qj,Witten:1998zw} that the Hawking-Page transition in
thermal AdS \cite{Hawking:1982dh}, which occurs at a critical
temperature, should be interpreted as a first order deconfinement
transition in field theory.

At weak coupling there are also known to be phase transitions as a
function of temperature
\cite{Sundborg:1999ue,Aharony:2003sx,Aharony:2005bq}. In the free
theory there is again a first order transition at a critical
temperature whilst at finite small coupling there are two
possibilities distinguished by the sign of a currently unknown
coefficient in the effective Lagrangian. Either there is a single
first order transition or, as one increases the temperature, a
second order transition followed by a third order transition.

The remaining question is the interpolation between weak and
strong coupling physics. A minimal interpolation suggests that the low
and high temperature regimes are separated by at least one phase
transition at all couplings \cite{Aharony:2003sx}.
Nonrenormalisation theorems in the low temperature phase
\cite{Brigante:2005bq} may suggest that the physics there depends
smoothly on the coupling. The high temperature behaviour is less clear.
In a recent work \cite{Hartnoll:2005ju} we emphasised the tension
between perturbative plasma physics correlators which generically
contain many threshold
branch cuts and gravitational computations of the same
correlators at strong coupling which generically contain poles in
frequency space.

The order parameter used to discuss these transitions is the
eigenvalue distribution of the Polyakov loop, the $SU(N)$ holonomy around
the Euclidean time circle
\be\label{eq:polyakov}
U = P\, e^{i \oint A_0 \, d\t} \,.
\ee
The eigenvalue distribution may be obtained from the expectation
values of the trace of powers of the Polyakov loop $\langle \Tr
U^k \rangle$. In the low temperature phase the distribution is
uniform while at high temperatures it is not. The possible third
order transition we mentioned is similar to the Gross-Witten
transition \cite{Aharony:2003sx,Gross:1980he,Alvarez-Gaume:2005fv},
in which the nonuniform eigenvalue distribution develops a gap.

The eigenvalue distribution of the Polyakov loop at strong
coupling is not known and it is the purpose of this work to
begin to approach the problem. An important question one would
like to address is whether the non-uniform high temperature
eigenvalue distribution at strong coupling is gapped or not. An
ungapped distribution would signal a third order phase transition
as a function of coupling in the high temperature phase
\cite{shenker}.

\subsection{Multiply wound Polyakov-Maldacena loops}

The AdS/CFT correspondence does not grant us access to the Polyakov
loop (\ref{eq:polyakov}) at strong coupling but instead allows us to compute the
Polyakov-Maldacena loop \cite{Maldacena:1998im,Rey:1998ik}
\be
U = P\, e^{i \oint [A_0 - i \Phi_I \theta^I(\t)] d\t} \,,
\ee
where $\Phi_I$ are the six adjoint scalar fields of the ${\mathcal{N}}=4$ theory
and $\theta^I(\t)$ is a trajectory on the unit $S^5$. This operator is
no longer unitary and so we cannot speak of an eigenvalue
distribution on a circle. Instead, the expectation values $\langle
\Tr U^k \rangle$ determine an eigenvalue distribution on the complex
plane which one can hope to use as an order
parameter. For instance, it is this Polyakov-Maldacena loop which
detects the strong coupling deconfinement transition.

At large $N$ and 't Hooft coupling $\lambda$ the AdS/CFT dictionary
gives \cite{Maldacena:1998im,Rey:1998ik,Gross:1998gk}
\be
\frac{1}{N} \langle \Tr U^k \rangle = e^{-S|_{k-winding}} \,,
\ee
where $S|_{k-winding}$ is the action of a classical string configuration
in the dual bulk geometry that winds the thermal circle $k$ times at the
conformal boundary. Most previous work, both for temporal and
spacelike Wilson loops, has focused on the single winding case
$k=1$. There are serious technical issues to overcome for
multiply wound strings.

A discussion of computing $\langle \Tr U^k \rangle$ at strong coupling
for spatial Wilson loops
was given in \cite{Gross:1998gk}. The approach is largely qualitative and
is essentially an extrapolation to four dimensions of intuition that
was developed in understanding two dimensional QCD as a simple string theory
\cite{Gross:1992tu,Gross:1993hu,Gross:1993yt}. The
classical string configuration is a disc worldsheet instanton
embedded in the dual geometry as a $k$ fold cover of a minimal
surface. The $k$ winding requires the insertion of $k-1$ $\bZ_2$
branch cuts running from interior points in the worldsheet to the
boundary. This suggests the following expression to leading order
in $\a'$ and string loop corrections
\cite{Gross:1998gk}
\be\label{eq:intuition}
\frac{1}{N} \langle \Tr U^k \rangle \sim (-1)^{k-1} A^{k-1} e^{- k S|_{k=1}} \,,
\ee
where the factor of $(-1)^{k-1}$ is due to the boundary conditions
of fermions around the loop and $A^{k-1}$ is a degeneracy factor
due to the different possible locations of the $k-1$ branch
points.

Despite the intuitive appeal of (\ref{eq:intuition}) it is not clear
what $A$ actually is. It could be the renormalised area of the
infinite worldsheet or for instance, as suggested in
\cite{Gross:1998gk}, it could be the area of
the worldsheet at the minimum of the background gravitational
potential. Doing the calculation rigorously requires computing the
measure on the moduli space of appropriate string instantons in the
dual background. This does not appear to be technically feasible at
the moment. Furthermore one might worry about the fact that the
worldsheets have conical singularities at the branch cuts.

In a beautiful paper, Drukker and Fiol have shown how to sidestep
these issues by using a single probe D3 brane carrying electric
worldvolume flux rather than fundamental strings
\cite{Drukker:2005kx}. In this approach there is no moduli space
and worldvolume curvatures are small everywhere. They computed the
value of $k$ winding supersymmetric circular Wilson loops and gave
a nontrivial match with a dual matrix model computation
\cite{Drukker:2000rr}. Even as $N \to \infty, \lambda \to \infty$
their result is significantly different from (\ref{eq:intuition})
if $\k = k \sqrt{\lambda}/4 N$ is kept fixed in this limit. In the
following sections we will adapt their method to compute $k$
winding Polyakov-Maldacena loops in the high temperature phase of
the theory. We will find however that for the Polyakov loop case,
nontrivial corrections come from D5 rather than D3 branes and
are expressed in terms of $\k' = k/N$.

Finally, we note that there is a perhaps underdeveloped parallel
between this critical string theory story and two dimensional
string theory on a circle. In that context the `Hawking-Page'
transition is seen as a Kosterlitz-Thouless transition in which
vortices condense on the string worldsheet \cite{Gross:1990md}.
This condensation precisely corresponds to the insertion of branch
cuts running to the boundary that we described above. Some of the
vortex condensates have been computed explicitly using the dual
large $N$ matrix quantum mechanics on a circle
\cite{Alexandrov:2001cm}.

In section 2 we consider potentially dual D3 branes in the black
hole geometry. We discuss possible boundary conditions and evaluation
of the action on the solutions. We find that the only solution to the
D3 brane equations of motion is a solution in which the D3 brane is
collapsed on the cigar submanifold of the black hole. These
configurations do not lead to the expected nontrivial $k$
dependence. In section 3 we move on to consider D5 branes.
Here we will find noncollapsed solutions that capture
the $k$ dependence of the dual Polyakov loop. The two main results of
this section are firstly the compution of the actions
with $\k'$ held fixed as $N \to \infty$ and secondly the observation
that there are interesting $N^{-2/3}$ corrections to the large $N$
limit with $k$ fixed. Section 4 is a summary and a discussion of
possible future directions.

\section{(No) D3 brane solutions}

In this section we search for probe D3 brane duals to the multiply
wound Polyakov loop. We will show that there are none, leading us on
to a study of D5 branes in the following section.

\subsection{Equations of motion}

The dual geometry for ${\mathcal{N}}=4$ SYM theory on $S^3 \times
S^1$ in the deconfined phase is the Euclidean Schwarzschild black
hole in $AdS_5
\times S^5$:
\be
ds^2 = R^2 \left[f(r) dt^2 + \frac{dr^2}{f(r)} + r^2 \left(d\a^2 +
  \sin^2\a d\Omega^2_2 \right) + d\Omega^2_5 \right] \,,
\ee
where
\be
f(r) = 1 - \frac{r_+^2 (1 + r_+^2)}{r^2} + r^2 \,.
\ee

Following \cite{Drukker:2005kx} we are looking for a probe D3
brane configuration in the dual background with the appropriate
symmetries and worldvolume flux to describe a multiply wound
Polyakov loop in field theory. The action for the probe D3 brane
is a sum of the Dirac-Born-Infeld and Wess-Zumino terms
\be\label{eq:action}
S = T_{D3} \int d\t d^3\sigma e^{-\Phi} \sqrt{\det \left( ^\star g
+ 2 \pi \a' F \right)} - i g_s T_{D3} \int {}^\star C_4 \,,
\ee
where the tension $T_{D3} = N/2\pi^2 R^4$. As usual $F$ is the
worldvolume gauge field and $^\star C_4$ is the pull back of the
bulk Ramond-Ramond four form potential. We can take the relevant
part of the potential to be
\be
C_4 = -\frac{i R^4}{g_s} r^4 \sin^2 \a dt \wedge d\alpha \wedge \text{vol} S^2 \,,
\ee
whilst the dilaton is contant: $e^{-\Phi} = 1$.

The probe brane configuration must have the same symmetries as the
dual Polyakov loop: $SO(3) \times SO(2) \subset SO(4) \times
SO(2)$, the isotropy group of a point times $S^1$ in $S^3 \times S^1$. If
$\{\alpha,\theta,\phi\}$ are coordinates on the $S^3$, then the
required configuration lies in the directions $t,\a,\theta,\phi,
r$, with a nontrivial dependence on only one worldvolume
coordinate
\be
\a = \a(\s) \,, \quad r = r(\s) \,.
\ee
The geometry of this embedding is the cigar of the Euclidean
black hole times an $S^2 \subset S^3$. Thus $\a(\sigma) \in [0,\pi]$
determines the size of the noncollapsed $S^2$ inside the $S^3$.
We can think of the D3
brane as $k$ fundamental strings blown up via the dielectric
Emparan-Myers effect \cite{Emparan:1997rt,Myers:1999ps}. The
string charge is induced from the worldvolume gauge field which
must have electric field strength with nonvanishing component
$F_{\t\s}(\s)$. The coefficient of the induced $B$ field coupling to
the string worldsheet will be the momentum $\delta S
/\delta F_{\t\s}$. This must be set equal to $i k$ to be
reinterpreted as the charge of $k$ fundamental strings. The presence
of $i$ will translate into the field strength being
imaginary as was found in \cite{Drukker:2005kx}. Therefore we introduce the
notation $F_{\t\s}(\s) \equiv i F(\s)$.
Evaluated on the ansatz we have just described, the action becomes
\be\label{eq:actionansatz}
S = \frac{2 N}{\pi} \int d\t d\s r^2 \sin^2\a
\left[\sqrt{\left(\frac{dr}{d\s}\right)^2 + r^2 f(r)
\left(\frac{d\a}{d\s}\right)^2 - \frac{4 \pi^2 F(\s)^2}{\lambda}} -
r^2 \frac{d\a}{d\s} \right]
\,,
\ee
where we used $R^4 = \lambda \a'^2$.

The equation of motion for the gauge field gives the total string charge $k
\in \bZ$
\be\label{eq:flux}
k = - \frac{\delta S}{\delta F} =
\frac{4N}{\lambda} \frac{2 \pi r^2 \sin^2\a
F}{\sqrt{\left(\frac{dr}{d\s}\right)^2 + r^2 f
\left(\frac{d\a}{d\s}\right)^2 - \frac{4 \pi^2
F^2}{\lambda}}} \,.
\ee
The equation of motion for $\a$ can be written
\be\label{eq:aeqn}
r^2 \sin^2\a\left[r^2 \sin 2\a \frac{2 \pi F}{\k \lambda^{1/2}} +
4 r \frac{dr}{d\sigma}\right] = \frac{d}{d\s}\left(r^2 f \frac{\k
\lambda^{1/2}}{2\pi F}\frac{d\a}{d\s}\right)\,,
\ee
where we introduced
\be
\k = \frac{k \sqrt{\lambda}}{4 N} \,.
\ee
The equation of
motion for $r$ is
\be\label{eq:reqn}
2 r^3 \sin^2\a \left[\sin^2\a \frac{2 \pi F}{\k \lambda^{1/2}} - 2
\frac{d\a}{d\s} \right] + \frac{1}{2} \frac{\k \lambda^{1/2}}{2 \pi
F} \frac{d(r^2 f)}{dr} \left(\frac{d\a}{d\sigma}\right)^2 =
\frac{d}{d\s}
\left(\frac{\k
\lambda^{1/2}}{2 \pi F} \frac{d r}{d \s} \right) \,.
\ee
One can use (\ref{eq:flux}) written as
\be\label{eq:feqn}
r^2 f \left(\frac{d\a}{d\s}\right)^2 +
\left(\frac{dr}{d\s}\right)^2 =  \frac{(2\pi F)^2}{\k^2
\lambda} \left(\k^2 + r^4 \sin^4\a \right) \,,
\ee
to eliminate $F$ from (\ref{eq:aeqn}) or (\ref{eq:reqn}). The
equations depend on $F$ only through the combination
\be
G = \frac{2 \pi F}{\k \lambda^{1/2}} \,,
\ee
so we will work in terms of this quantity in what follows.

Reparametrisation invariance of the action suggests that we should
be able to consistently set
\be
r = \sigma \quad \text{or} \quad \a = \sigma \,,
\ee
and then solve for $\a(r)$ or $r(\a)$, respectively. Using the
three equations of motion above it is straightforward to show that
indeed we may do this.

\subsection{Boundary terms and boundary conditions}

Once we have solved the equations of motion, we need to evaluate
the action on the solution to obtain the expectation value of the
$k$ winding Polyakov-Maldacena loop
\be
\langle \Tr U^k \rangle = e^{-S|_{k-soln}}.
\ee
As is discussed very clearly in
\cite{Drukker:2005kx}, in evaluating the action on the solution it
is important to add the boundary terms. These terms implement
Legendre transformations that permit the solution to have the
correct boundary conditions: fixed winding number $k$ and fixed
momentum in the $r$ direction.

For our configurations we need to add
\be
\left. S \right|_{bdy.} = - \int d\t \left. r \frac{\d S}{\d \pa_\s r} \right|_{r \to
\infty}  - \int d\t \left. A_t \frac{\d S}{\d \pa_\s A_t}\right|^{r \to \infty}_{r = r_+}
\,.
\ee
Using the action (\ref{eq:action}) this term becomes
\be
\left. S \right|_{bdy.} = - \frac{2 N}{\pi} \int \left. d\t \frac{r}{G}
\frac{dr}{d\sigma}\right|_{r \to \infty}
+ \frac{2 N}{\pi} \int_{r_+}^{\infty} d\t d\s \k^2 G \,.
\ee
In the above expressions, and in this section in general, we have
assumed for simplicity that the configuration reaches the horizon
$r_+$. However it is possible for the solution to close off at a
finite value $r_\text{min} > r_+$ by having $\a(r_\text{min})=0$. We
will consider this possibility in later sections.

There is a further ambiguity in the boundary term following from
the fact that large gauge transformations of the background four
form potential can change the Wess-Zumino part of the action.
However, it was found in \cite{Drukker:2005kx} that using a
minimal expression for the potential compatible with the
symmetries of the problem, as we have done, together with no extra
boundary term gives the correct answer for cases that can be
matched.

In order for a D3 brane configuration to contribute
semiclassically to a dual Polyakov loop it must have finite
action, including the bulk and boundary terms. Let us look at the
large $r$ behaviour of solutions. We have found two asymptotic
behaviours for the fields $\a(r)$ and $G(r)$ at large $r$ that
lead to a finite total action
\bea\label{eq:rlarge}
\k G(r) & = & 1 + \frac{1}{2} \frac{\k^2}{r^2} - \frac{4 B}{3} \frac{\k^3}{r^3} +
\cdots \,, \nonumber \\
\a(r) & = & \pi - \frac{\k}{r} + B \frac{\k^2}{r^2} - \frac{1 + 6
  B^2}{6}\frac{\k^3}{r^3} + \cdots \,,
\eea
where $B$ is an undetermined constant, and
\bea\label{eq:rlarge3}
\k G(r) & = & 1 + \frac{9 A^2}{2} \frac{\k^4}{r^4} - \frac{3 A^2 (4 A \k
  + 3)}{2 \k^2} \frac{\k^6}{r^6} + \cdots \,, \nonumber \\
\a(r) & = & \pi - \frac{A}{\k} \frac{\k^3}{r^3} + \frac{A(2 A \k + 3)}{5
  \k^3} \frac{\k^5}{r^5} + \cdots\,,
\eea
where $A$ is a free constant. A dependence on the horizon size
$r_+$ enters at higher orders in these expansions. We have written
the expansions where $\a \to \pi$ as $r \to \infty$. There are
very similar expansions for $\a \to 0$. The $\a \to \pi$ case is
more relevant as configurations with increasing $\a$ have a lower
action because of the Wess-Zumino term in (\ref{eq:actionansatz}).
In fact this term causes the action to behave very differently in
the cases $\a \to \pi$ from below and $\a \to 0$ from above.
Whilst both (\ref{eq:rlarge}) and (\ref{eq:rlarge3}) lead to
finite action as $\a \to \pi$, only the faster falloff
(\ref{eq:rlarge3}) is admissible as $\a \to 0$.

The falloffs (\ref{eq:rlarge}) and (\ref{eq:rlarge3}) are special
one parameter subfamilies of the general behaviour of solutions
near infinity, which depends on two constants of integration. The
generic behaviour at large $r$ is
\bea\label{eq:generic}
G(r) & = & \frac{m(r) - r m'(r)}{m(r)^2} + \cdots \,, \nonumber
\\
\a(r) & = & \pi - \frac{m(r)}{r} + \cdots \,,
\eea
where $m(r) \to \infty$ as $r \to \infty$ but $m(r)/r
\to 0$. The explicit leading order behaviour of $m(r)$ appears to
be somewhat complicated. Numerically we find that $m(r)
\sim \log r$ often gives a good description, but is not exact.
However, without solving the equations for $m(r)$ it is possible
to show that the action of these solutions behaves for large $r$
as
\be
\left. S \right|_{soln.} = \frac{-2 N \b}{\pi} \int^{\infty}
\frac{dr}{2} \frac{m(r)^2}{m(r) - r m'(r)} + \cdots \to - \infty \,.
\ee
This divergence may be very precisely checked numerically.
Therefore configurations with the generic behaviour
(\ref{eq:generic}) have infinite action and do not contribute to
the Polyakov loop computation. It is not that the action is
becoming unbounded below but rather that these boundary conditions
do not give rise to any normalisable states.

The bulk action is in fact finite with the asymptotic behaviours
(\ref{eq:rlarge}) and (\ref{eq:rlarge3}) with another finite
contribution to the action coming from the combined boundary
terms. The action evaluated on a solution with all boundary terms
included may be written
\bea
\left. S \right|_{soln.} & = & \left. \frac{2 N \b}{\pi} \right( - r_+ \sqrt{\k^2
  + r_+^4 \sin^4\a_+} \nonumber \\
 & & +
\left. \int_{r_+}^{\infty} \left[\frac{r^2 f(r)}{G} \left(\frac{d\a}{dr}\right)^2
- r^4 \sin^2\a \frac{d\a}{dr} + \frac{r}{G^2} \frac{d G}{dr}
\right] dr \right)
\eea
where
\be
\b = \frac{2\pi r_+}{1+2 r_+^2} \,,
\ee
is the period of the time circle. The simplest solution is the
collapsed solution which has $\a = 0$ or $\a = \pi$. In this case
the action is given immediately as
\be\label{eq:collapsedaction}
\left. S \right|_{collapsed} = - \frac{2 N \k \b r_+}{\pi} =
- \frac{\sqrt{\lambda} k \b r_+}{2\pi} \,.
\ee
This is $k$ times the action for a fundamental string instanton
wrapping the cigar. This connection between
collapsed D branes carrying electric flux and fundamental strings
has been consistently verified in a range of examples since
\cite{Emparan:1997rt}.

At this point we should note that for a general solution the
dependence of the Polyakov loop on $k$, $N$ and $\lambda$ is
significantly restricted to be of the form
\be\label{eq:generalform}
\left. S \right|_{soln.} = N \b s(\k,r_+) \,,
\ee
for some function $s(\k,r_+)$. As was emphasised in
\cite{Drukker:2005kx}, although the expression
(\ref{eq:generalform}) is leading order in $1/N$ and $\lambda$, it
is exact in $\k = k \sqrt{\lambda} / 4 N$. This represents a
partial resumation of bulk higher genus string corrections. Such
$\k$ dependence could not have been seen in the heuristic
arguments of \cite{Gross:1998gk} leading to (\ref{eq:intuition})
that we reviewed in the
introduction. The result of \cite{Drukker:2005kx} for the $k$
winding circular spatial Wilson loop, $S_k = - 2 N \left[\k
\sqrt{1+\k^2}+\sinh^{-1} \k \right]$, reduces to the
na\"ive expectation of $S_k = k S_1$ only in the limit $\k
\to 0$.

The important lesson from \cite{Drukker:2005kx} is therefore that
the possibility of a D3 brane blowing up due to an Emparan-Myers
effect corresponds to and captures the nontrivial $\lambda$ and $N$
corrections which one expects from the singular nature of the
fundamental string picture. These corrections are contained in the
potentially nontrivial dependence of (\ref{eq:generalform}) on $\k = k
\sqrt{\lambda}/4N$. We will now see that in contrast to the spatial
Wilson loop case, this potential is not realised for the Polyakov
loop. Instead, we will need to consider D5 branes in the following
section and find that these do give corrections to the collapsed result
in terms of $\k' = k/N$. At large $\lambda$ these are subleading
compared with the $\k$ corrections which appear to vanish.

\subsection{Solving the equations}

After extensive numerical investigation of the equations of motion
(\ref{eq:aeqn}), (\ref{eq:reqn}) and (\ref{eq:feqn}), it has
become clear that the only solutions which run from infinity to
the horizon are the collapsed solutions $\a = 0$ and $\a = \pi$,
with action (\ref{eq:collapsedaction}). In this section we will combine
analytic and numerical arguments to understand this fact.

We begin with some analytic results that are possible at large
values of $\k$. We will see in this regime that noncollapsed
solutions do not reach the horizon but rather turn around at some
minimum radius $r_*$ and then go back out to infinity. We will
then review continuity theorems for the solutions as a function of
$\k$ and the boundary conditions to show that this behaviour
persists for a range of solutions away from the large $\k$ limit.
Finally, we present numerical data indicating that none of the
noncollapsed solutions reach the horizon or close off in the
interior.

We will jump between considering $r(\a)$ and
$\a(r)$ depending on what is most convenient. $G(\a)$ or $G(r)$
can be completely eliminated from the problem so we will not
discuss its behaviour explicitly.

\subsubsection{The solutions at large $\k$}

The equations of motion simplify when $\k$ is taken to be large.
One may find the general solutions that have asymptotic boundary
conditions (\ref{eq:rlarge}) and (\ref{eq:rlarge3}). With the
former boundary condition the solution is
\be\label{eq:sol1}
r(\a) = \k \frac{1 + B (\a-\pi)}{\sin\a} \,,
\ee
which can of course be viewed as a transcendental equation for
$\a(r)$. In the latter case we have
\be\label{eq:sol2}
\a(r) = \pi - \frac{3A}{\k} \int_{r/\k}^{\infty} \frac{dy}{y^2 \sqrt{y^4 - 9 A^2}}\,.
\ee
The large $\k$ expansion is valid for both of these solutions
provided the constants $B$ and $A$ are of order one and for the
coordinate range $r_+ \ll r \ll \k^3$. Both the expressions
(\ref{eq:sol1}) and (\ref{eq:sol2}) agree excellently with
numerics.

If we think of these solutions as $r(\a)$, then we see that as the
solution comes in towards the horizon from $r \to \infty$ it turns
around at some minimal radius $r_* \sim \k \gg r_+$. The behaviour
near the turnaround point is
\be\label{eq:turnaround}
r(\a) = r_* + C^2 \left(\a - \a_* \right)^2 + \cdots \,.
\ee
The values of the constants $r_*, \a_*$ and $C$ may be determined
as functions of $B$ or $A$ from the solutions (\ref{eq:sol1}) and
(\ref{eq:sol2}), respectively. If $A$ and $B$ are order one, then
the turnaround occurs within the regime of validity of the
solution. We note however that (\ref{eq:sol2}) will only be valid
for the branch of the solution going from $\a=\pi$ to $\a=\a_*$.

None of these noncollapsed solutions reach the horizon and close
off. Instead they come in from infinity at $\a = \pi$ and turn
around and go back out to infinity at $\a = 0$. The solutions
are illustrated schematically in Figure 1. The
interpretation of these solutions is as potential contributors to the
two point function
of the Polyakov loop at antipodal points of the field theory
$S^3$, that is $\langle \tr U^{\dagger k}(0)
\tr U^k(\pi)\rangle$. However all of these solutions have an
infinite action due to their behaviour as $\a \to 0$.
It is likely that the dual
configuration for this antipodal two point function will be the
disconnected sum of the one point functions, corresponding to a
total screening of external quarks. This phenomenon is known
to occur for the singly wound case
\cite{Brandhuber:1998bs, Rey:1998bq, Landsteiner:1999up}.

\begin{figure}[h]
\begin{center}
\epsfig{file=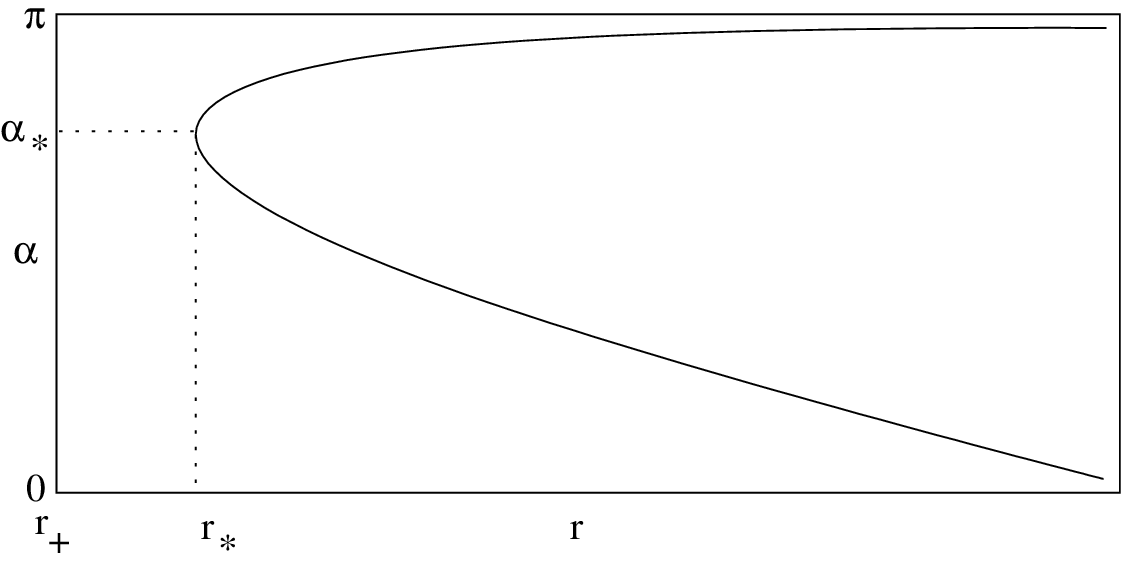,width=8cm}
\end{center}

\noindent {\bf Figure 1:} The solution reaches down to a minimal
radius $r_* > r_+$ and then goes back out to infinity.

\end{figure}

Thus, at large $\k$ at least and provided we don't take $A$ or $B$
to be very small, the only solution which could contribute to the
expectation value of the $k$ winding Polyakov loop is the
collapsed solution. We will now argue that this continues to be
true for all values of $\k$ and all values of $A$ and $B$.

\subsubsection{Continuity as a function of $\k$ and boundary conditions}

To study the existence and properties of solutions to ordinary
differential equations, the key question to establish is where the
equation satisfies the Lipshitz property. To do this we write the
equation for $\a(r)$ as two first order equations
\bea\label{eq:firstorder}
\frac{d\a}{dr} & = & \g \,, \nonumber \\
\frac{d\g}{dr} & = & F(\a,\g,r) \,,
\eea
where the explicit expression for $F(\a,\g,r)$ is easily
computable and not particularly illuminating. The Lipshitz
property in this case boils down to requiring that for all
$t,\a_1,\a_2,\g_1,\g_2$ in some rectangle we have
\be
[F(\a_1,\g_1,r)-F(\a_1,\g_1,r)]^2 \leq L \left( [\a_1-\a_2]^2 +
[\g_1 - \g_2]^2 \right)
\ee
for some constant $L$. Where the Lipshitz property holds,
solutions to the equations exist, are unique given initial values
of $\a$ and $\g$, and depend continuously on initial conditions
and on parameters in the equation \cite{ode}.

For our equation (\ref{eq:firstorder}) the Lipshitz property may
be seen to hold everywhere except when $r \to r_+$, $r \to
\infty$ or $\g \to \infty$. Therefore, away from infinity and the
horizon, the only way the solution can terminate at some finite
$r_*$ is if the derivative $\a'(r)$ diverges at that point. We
have seen that indeed this occurs in the large $\k$ solutions, as
(\ref{eq:turnaround}) implies $\a \sim (r-r_*)^{1/2}$. One may
show that such square root behaviour is the only allowed behaviour
as $\a'(r) \to \infty$.

Continuity then implies that as we vary $\k$ and the constants
$A,B$ in the finite action boundary conditions (\ref{eq:rlarge})
and (\ref{eq:rlarge3}), all the solutions coming in from infinity
will turn around at some finite $r_*$, so long as $r_*$ does not
approach the horizon $r_+$. As we approach $r_+$ the continuity
argument no longer holds. Another possible termination is that
$\a(r_\text{min}) = 0$ for some $r_\text{min} > r_+$ which would
allow us to truncate the solution at that point as the $S^2$ closes off
there.

Although continuity therefore gives us a window around the large
$\k$ solutions in which the behaviour is qualitatively similar, it
is not enough to cover all possible $\k$ and initial data. To this
end we now present some numerical results.

\subsubsection{The minimum radius as a function of boundary conditions}

Figure 2 shows how the angle $\a_*$ at which the turnaround occurs
varies as the minimum radius $r_*$ comes close to the horizon
$r_+$. The plot is obtained by varying the boundary conditions at
infinity. A solution will reach the horizon if $r_* \to r_+$. However,
what we see happenning is that as $r_* \to r_+$ then $\a_* \to \pi$.
That is, as the solution gets closer and closer to the horizon, it
tends towards the collapsed solution. The
behaviour is illustrated in Figure 2 for three values of the horizon
size, but in fact occurs for all horizon sizes from thermal AdS,
which has $r_+ = 0$, to the large horizon limit $r_+ \to \infty$, which
may be described using a planar horizon.

\begin{figure}[h]
\begin{center}
\epsfig{file=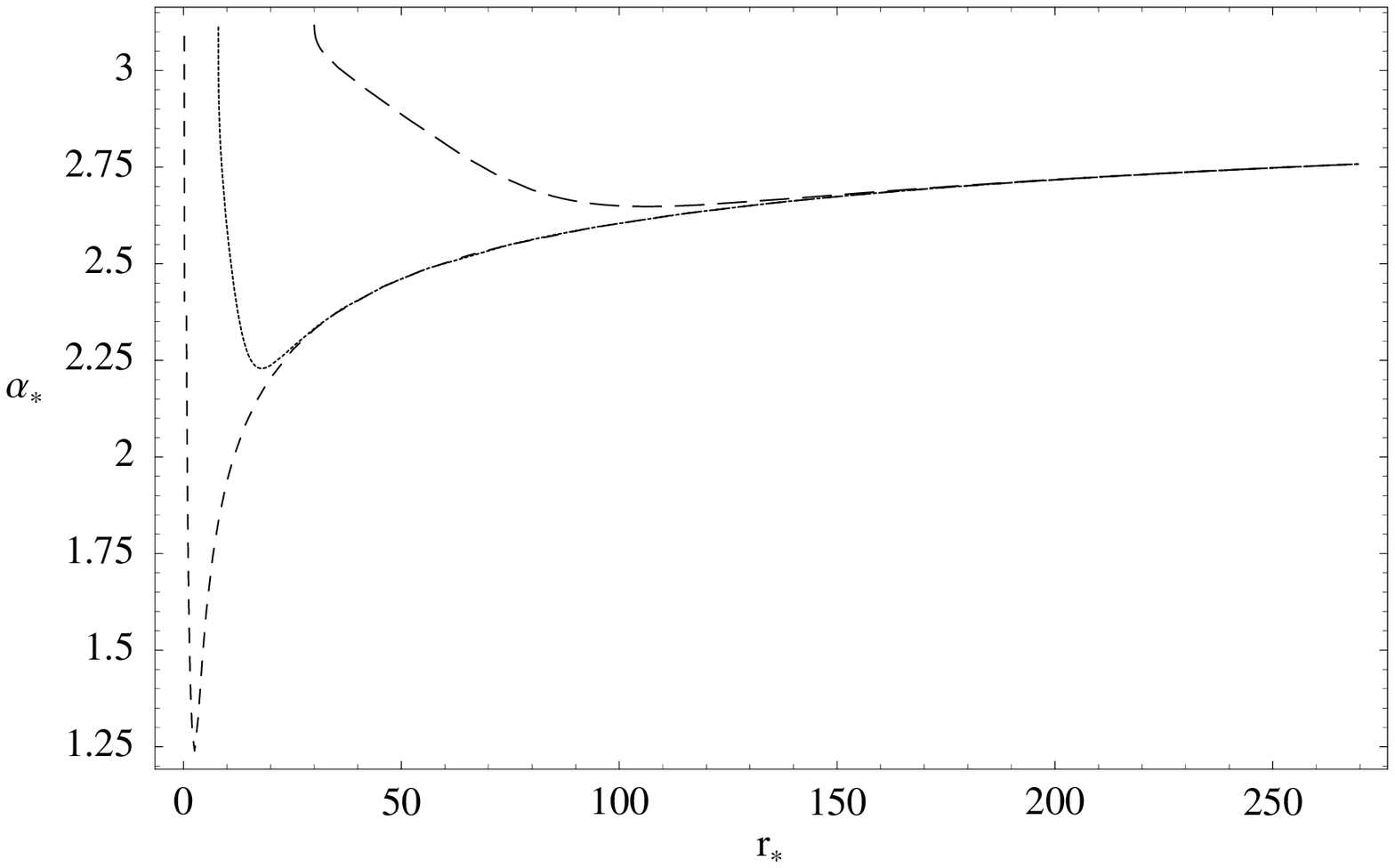,width=10cm}
\end{center}

\noindent {\bf Figure 2:} The value of $\a_*$ at which the turnaround radius
$r_*$ is reached for three values of the horizon size $r_+$. From
bottom to top $r_+ = 0.2, 8, 30$. All plots have $\k=5$ and asymptotic
behaviour (\ref{eq:rlarge}). We find that $\a_* \to \pi$ as $r_* \to
r_+$.

\end{figure}

In Figure 2 we have used solutions that have the falloff
(\ref{eq:rlarge}) as $r \to \infty$. We also performed the
computation with the faster falloff (\ref{eq:rlarge3}) and found
exactly the same behaviour. As a crosscheck of these results, we also
integrated numerically out from the horizon. We found that the
solutions always reach $r \to \infty$ with a non normalisable
falloff. The behaviour at infinity tends towards the normalisable
falloff as $\a(r_+) \to \pi$, in perfect agreement with the results
we have just described for integrating in.

Therefore, we seem to find strong numerical evidence that all noncollapsed
solutions satisying normalisable boundary conditions as $r \to
\infty$ do not reach the horizon. We have also seen that they do not close
off by reaching $\a = 0$ at some $r_\text{min}$. It follows that the
only D3 brane configurations that could contribute to the dual multiply
wound Polyakov loop expectation value are the collapsed solutions.

We also considered whether any of the noncollapsed solutions gave a
finite action contribution to the two point function $\langle \tr
U^{\dagger k}(0) \tr U^k(\pi)\rangle$. It seems that none of the
solutions have normalisable falloffs at both $\a \to \pi$ and $\a \to
0$.

\subsubsection{Implications of the D3 brane results}

In this section we have studied D3 brane embeddings into the
Schwarzschild AdS background, searching for dual configurations to
multiply wound Polyakov-Maldacena loops in the strongly coupled
large $N$ thermal field theory. The impressive
matching of \cite{Drukker:2005kx}, for the case
of multiply wound spatial Wilson loops, indicated that the blowing up
of an $S^2$ in the D3 brane worldvolume corresponds to a departure
from the na\"ive expectation $S_k =k S_1$ for the action of a multiply
wound string worldsheet. This implied that for spatial Wilson loops
$\frac{1}{N} \langle\Tr U^k\rangle\neq \frac{1}{N^k} \langle\Tr
U\rangle^k$, generically. We have found that
this blow up does not occur in the case of the Polyakov-Maldacena loops
at strong coupling. If this were the full story it would imply
instead $\frac{1}{N} \langle\Tr U^k\rangle =
\frac{1}{N^k} \langle\Tr U\rangle^k$ following directly from
(\ref{eq:collapsedaction}).

As we have discussed above, the Polyakov-Maldacena loop differs from
the Polyakov loop by the presence of scalar fields and is not
unitary. A priori one might expect $N^2$ degrees of freedom and hence
obtain an eigenvalue distribution on a plane rather than a
circle. However, the $\frac{1}{N} \langle\Tr U^k \rangle$ can be used to define
eigenvalue distributions on a circle. Writing $z = r e^{i \theta}$ and
fixing the modulus $r$ we can obtain a distribution on the $\theta$ circle
\be
\rho_r(\theta) = 1 + 2 \sum_{k=1}^{\infty} r^k \cos k \theta
\frac{1}{N} \langle\Tr U^k \rangle \,.
\ee
When $\frac{1}{N} \langle\Tr U^k \rangle=
\frac{1}{N^k} \langle\Tr U\rangle^k$ there are three possibilities,
depending on the value of $a = r \frac{1}{N} \langle\Tr U \rangle$.
If $a = 1$ then the distribution is a delta function. If $a < 1$ then
the distribution becomes
\be\label{eq:converge}
\rho_r(\theta) = \frac{1 - a^2}{1 + a^2 - 2 a \cos \theta} \,.
\ee 
This is a smooth ungapped distribution on the $\theta$ circle.
If $a > 1$ then the sum does not converge. Analytic continuation in
$a$ leads to the result (\ref{eq:converge}), but now the distribution
has a singularity that is not integrable.

It remains to be seen what the precise relation of these distributions
is to the eigenvalue distribution of $U$, and whether they can be used
as order parameters. Furthermore,
the result $\frac{1}{N} \langle\Tr U^k\rangle =
\frac{1}{N^k} \langle \Tr
U\rangle^k$ is clearly not reliable. The initial motivation for
studying D3 branes was precisely to avoid issues that arise in the
multiply wound string picture. However, the fact that the D3 brane
configurations are collapsed seems to bring us back to that
picture. Although the collapsed D3 brane retains a finite tension
which is exactly $k$ times the fundamental string tension, it
remains true that induced curvatures on the collapsed $S^2$ are large
and not controlled within the validity of the Dirac-Born-Infeld action.

If the corrections to the multiply wound string picture are necessarily
captured by a dependence on $\k = k \sqrt{\lambda}/4N$ then the fact
that we do not find these corrections would be sufficient to imply that the
eigenvalue distributions discussed above are indeed possible at leading
order in $N$ and $\lambda$. However, we will see in the following
section that by considering D5 brane probes instead of D3 branes
one finds corrections in terms of $\k' = k/N$. Note that in the large
$\lambda$ limit $\k' \ll \k$ and so these corrections are subleading
with respect to potential D3 brane corrections.

\section{D5 brane solutions}

In this section we search for probe D5 brane configurations in the
Euclidean AdS Schwarzschild background that have the correct charge
and symmetries to contribute to multiply wound Polyakov loops. The
logic will be close to that of the previous section so we shall be
briefer in our presentation.

\subsection{Equations of motion}

The action for the probe D5 brane is again a sum of Dirac-Born-Infeld
and Wess-Zumino terms
\be
S = T_{D5} \int d\t d^5\sigma e^{-\Phi} \sqrt{\det \left( ^\star g
+ 2 \pi \a' F \right)} - i g_s T_{D5} \int 2 \pi \a' F \wedge
{}^\star C_4 \,,
\ee
where $T_{D5} = N \sqrt{\lambda}/8 \pi^4 R^6$. As we will be
considering solutions that are blown up in the $S^5$ direction, the
relevant part of the four form potential is now
\be
C_4 = \frac{R^4}{g_s} \left[\frac{3 (\g-\pi)}{2} - \sin^3\g \cos\g -
  \frac{3}{2} \cos\g \sin\g \right] \text{vol} S^4 \,,
\ee
where $\g \in [0,\pi]$ is a polar coordinate on the $S^5$.

We need a configuration that has the symmetries of the dual Polyakov
loop: the isotropy group $SO(3) \times SO(2)$ of a point times $S^1$
in $S^3 \times S^1$, as well as an $SO(5) \subset SO(6)$ of the R symmetry group.
This is obtained by having the D6 brane wrap an $S^4$ in the $S^5$ and
wrapping the time circle, while remaining at a point in the horizon
$S^3$. Thus again the only nontrivial dependence is in $\g(\sigma)$
and $r(\sigma)$. The electric worldvolume gauge field will again be
imaginary, so as before we let $F_{\t\s}(\s) \equiv i
F(\s)$. Evaluated on this ansatz, the action becomes
\be
S = \frac{N \sqrt{\lambda}}{3 \pi^2} \int d\t d\sigma \left[ \sin^4\g
  \sqrt{\left(\frac{dr}{d\s} \right)^2 + f(r) \left(\frac{d\g}{d\s}
    \right)^2 - \frac{4 \pi^2 F(\s)^2}{\lambda}} -
    D(\g) \frac{2 \pi F(\s)}{\lambda^{1/2}} \right] \,,
\ee
where we used $R^4 = \lambda \a'^2$ and introduced
\be
D(\g) = \sin^3\g \cos\g + \frac{3}{2} \cos\g \sin\g - \frac{3 (\g-\pi)}{2}
\,.
\ee

We will set $r = \s$ and look for solutions for $\g(r)$. The
equation of motion for $F$ is
\be
k = - \frac{\d S}{\d F} = \frac{2 N}{3 \pi} \left[\frac{2 \pi
    F}{\lambda^{1/2}} \frac{\sin^4\g}{\sqrt{1 + f(r) \left(\frac{d\g}{dr}
    \right)^2 - \frac{4 \pi^2 F^2}{\lambda}}} + D(\g) \right] \,,
\ee
where as before $k \in \bZ$ is the induced fundamental string
charge. The equation of motion for $\g$ is
\be
 4 \sin^4\g \left[ \frac{\sin^3\g \cos\g}{\frac{3\pi \k'}{2} - D(\g)}
+ 1 \right] \frac{2 \pi F}{\lambda^{1/2}} =
\frac{d}{dr} \left(\frac{\lambda^{1/2}}{2 \pi F}\left[\frac{3\pi
    \k'}{2} - D(\g) \right] f(r) \frac{d\g}{dr} \right)\,,
\ee
where we introduced
\be
\k' = \frac{k}{N} \,.
\ee
It will also be convenient to introduce
\be
G = \frac{2 \pi F}{\lambda^{1/2}} \,.
\ee
Note that the equations of motion are invariant under $k \to N - k$
and $\g \to \pi - \g$, reflecting the fact that the Polyakov loop is
sensitive to the N-ality of the source.

These equations are closely related to the D5 brane configuration that
is dual to the baryon vertex \cite{Callan:1998iq, Callan:1999zf,
Gomis:2002xg}. In
fact, the solutions we will present clarify the uncertainty in
these papers concerning the interpretation of solutions with
$k < N$ units of worldvolume flux. The simplest solutions we find
will be at a constant angle $\g_0$ on the $S^5$, similar
in some regards to the confining string solutions described in
\cite{Herzog:2001fq}. The connection between the baryon vertex and
confining strings was realised particularly explicitly in
\cite{Hartnoll:2004yr}.

\subsection{Boundary terms and action}

The boundary terms to be added to the action are similar to the D3
brane case. The action now depends on the derivative of one of
the coordinates on the $S^5$, $\g$. Therefore we need to include an extra
boundary term to impose Neumann boundary conditions in this
direction. A clear discussion of these issues can be found in
\cite{Drukker:2005kx}. The full boundary term we need to add can be
written as
\bea
\left. S \right|_{bdy.} & = &
- \frac{N \sqrt{\lambda}}{3 \pi^2} \left. \int d\t \frac{1}{G} \left[\frac{3 \pi \k'}{2} -
D(\g) \right] \left[r + (\g-\pi) f(r) \frac{d\g}{dr} \right] \right|_{r \to
  \infty} \nonumber \\
 & & + \frac{N \sqrt{\lambda}}{3 \pi^2} \int_{r_+}^{\infty} d\t dr
\frac{3 \pi \k'}{2} G \,.
\eea

As with the D3 branes, there are collapsed
solutions with $\g = \pi$. As in the D3 brane case, these solutions
take us back to the multiply wound string picture and are not
reliable. These are again seen to have
action
\be
S|_{collapsed} = - \frac{\sqrt{\lambda} k \b r_+}{2\pi} \,,
\ee
which is $k$ times the action for a fundamental string instanton
wrapping the cigar. For a general solution we will have
\be
S|_{soln.} = N \sqrt{\lambda} \b s(\k',r_+) \,,
\ee
for some function $s(\k',r_+)$. The D5 brane probes are sensitive to
different corrections to the dual Polyakov loop than the D3 branes,
depending on $\k'$ rather than $\k$. Note that $\k' \ll \k$ in the
large $\lambda$ limit.

As $r \to \infty$, the following falloff is allowed by the equations
of motion and leads to a finite action configuration:
\be
\g(r) = \pi - \frac{A}{r} + \frac{A (2 - A^2)}{6} \frac{1}{r^3} +
\frac{2 A^4}{9 \pi} \frac{1}{\k'} \frac{1}{r^4} + \cdots \,,
\ee
with $A$ an arbitrary constant. This is the falloff considered by
\cite{Callan:1998iq, Callan:1999zf} and, if the time circle were not
thermal, would give an asymptotically supersymmetric
configuration.

There is another type of finite action solution which
has qualitatively different behaviour. These are solutions in which
the angle $\g$ tends to a constant $\g_0$, determined by $\k'$, as $r
\to \infty$. In particular, these include configurations in which $\g
= \g_0$ is constant everywhere.

In contrast to the D3 brane case, numerical investigation of the
equations reveals that integrating inwards with these boundary
conditions leads to solutions that reach the horizon or close off at
some radius where $\g(r_\text{min})=0$. Some possibilities are
illustrated in figure 3. The case where $\g(r_\text{min})=0$ is was
considered by \cite{Callan:1998iq, Callan:1999zf}. In this case the
solution is in the same homology class as the $S^5$
which therefore induces $N$ units of flux on the
worldvolume. These solutions are only possible with $k = N$
and correspond to baryon vertices in field theory. We are more
interested in the solutions which reach the horizon as these can have
arbitrary $k$ and can therefore contribute to multiply wound Polyakov
loops. These solutions can only exist because of the presence of a
horizon, which corresponds to being in the deconfined phase in which
nonsinglets can be screened.

\begin{figure}[h]
\begin{center}
\epsfig{file=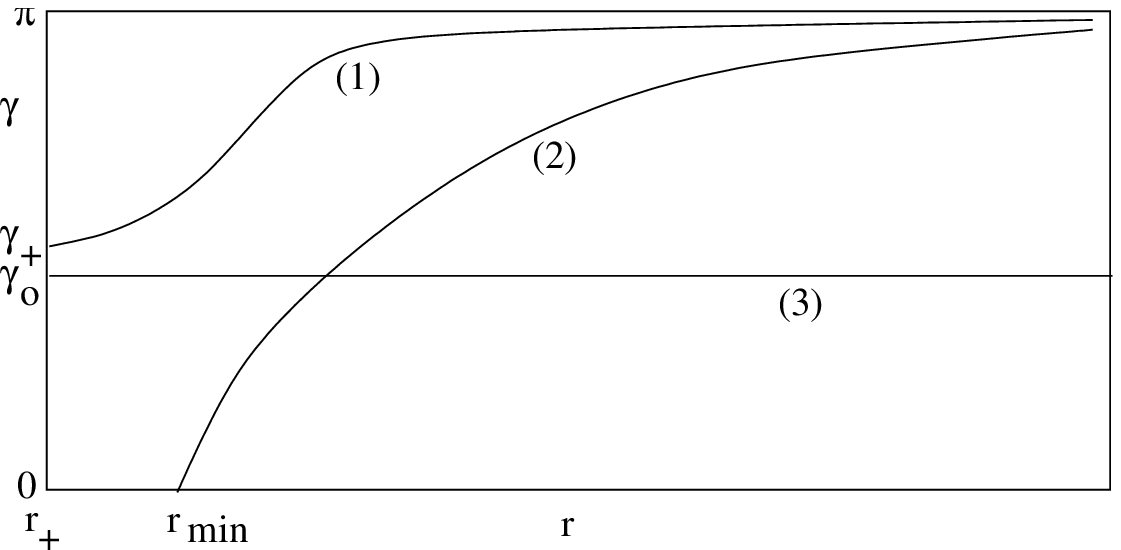,width=8cm}
\end{center}

\noindent {\bf Figure 3:} Three types of D5 brane solutions. In (1)
the solution runs from $\g = \pi$ to $\g = \g_+$ at the horizon. In
(2) the solution runs from $\g = \pi$ to $\g = 0$. In (3) the solution
is at a constant $\g = \g_0$.

\end{figure}

We postpone a systematic study of solutions to the D5 brane equations,
and their field theory interpretation, for future work.
Instead, in the following two subsections we study firstly
the simplest solutions, which are constant $\g = \g_0$, and secondly
we look at the $\k' \to 0$ limit which is the most relevant for making
direct contact with dual field theory computations.

\subsection{Constant solutions}

The configuration $\g = \g_0$ is a solution to the equations of motion
if $\g_0$ satisfies
\be\label{eq:constant}
\pi (\k'-1) =  \sin\g_0 \cos\g_0 - \g_0 \,.
\ee
This remarkable solution is possible because the metric function
$f(r)$ only appears in the equations of motion multiplied by
$d\g/dr$, and so drops out when $\g$ is constant. There is a unique
value of $\g_0 \in [0,\pi]$ for each value of $\k' \in [0,1]$.

The action, with boundary terms included, evaluated on these solutions
is finite and may be written
\be\label{eq:constaction}
S|_{\g_0} = - \frac{N \sqrt{\lambda} \b r_+}{3 \pi^2} \sin^3\g_0 \,.
\ee
The action (\ref{eq:constaction}) is exact if $\k'$ is held fixed in
the $N \to \infty$ limit. This is not usually what one does in field
theory, where the large $N$ limit is taken prior to computing Wilson
loop observables, with the notable exception of the supersymmetric
circular Wilson loop in ${\mathcal{N}}=4$ SYM theory
which allows subleading $N$ corrections to be
computed using a matrix model \cite{Drukker:2000rr,Drukker:2005kx}.
We will consider the more usual $N \to
\infty$ limit in the following subsection.
It is also not clear that these constant solutions are the
correct duals for $\frac{1}{N} \langle \tr U^k \rangle$, or any other
higher loop in a representation of N-ality $k$, as opposed to
the non constant solutions discussed around figure 3. However, given that
the constant solutions may be characterised so explicitly, let us
consider the eigenvalue distribution that would follow from
using (\ref{eq:constaction}) to compute $\frac{1}{N} \langle \tr U^k
\rangle$. As we note below, recent developments disfavour this
interpretation. However, given the unexpectedly interesting result
shown in figure 4, it seems possible that this computation will play a
role in a more fully developed understanding.

We will compute the eigenvalues numerically for a fixed, relatively
large value of $N$. If we were to assume that the result
(\ref{eq:constaction}) gave us the
traces
\be\label{eq:consttraces}
\frac{1}{N} \langle \Tr U^k \rangle = e^{N \frac{2}{3\pi} \eta
  \sin^3\g_0(k)} \,,
\ee
then we may determine the eigenvalues by solving a system of
equations. Here we introduced
\be\label{eq:eta}
\eta = \frac{\sqrt{\lambda} \beta r_+}{2\pi} \,,
\ee
which is closely related to the action for a single fundamental
string. We should take $\eta \gg 1$ corresponding to large 't Hooft
coupling. At high temperatures $\b r_+ \sim
{\mathcal{O}}(1)$. Unfortunately, this makes the system of equations
which determine the eigenvalues numerically rather intractible because
the presence of $N \eta$ in the exponent means that some equations
have extremely large entries and others have entries of order one. We
will not solve this problem here. Instead, let us take $\eta \sim
1/N \ll 1$. This of course takes us completely outside the regime of
validity of the derivation of (\ref{eq:constaction})
and leads to the important question of how strongly the eigenvalue
distribution depends on $\eta$. Nonetheless, one obtains an interesting
result. Figure 4 shows the eigenvalue distribution following from
(\ref{eq:consttraces}) with $N=35$. The eigenvalues lie on
a circle in the complex plane, fully to the right of the imaginary
axis. They are continuously distributed,
with no gaps.

\begin{figure}[h]
\begin{center}
\epsfig{file=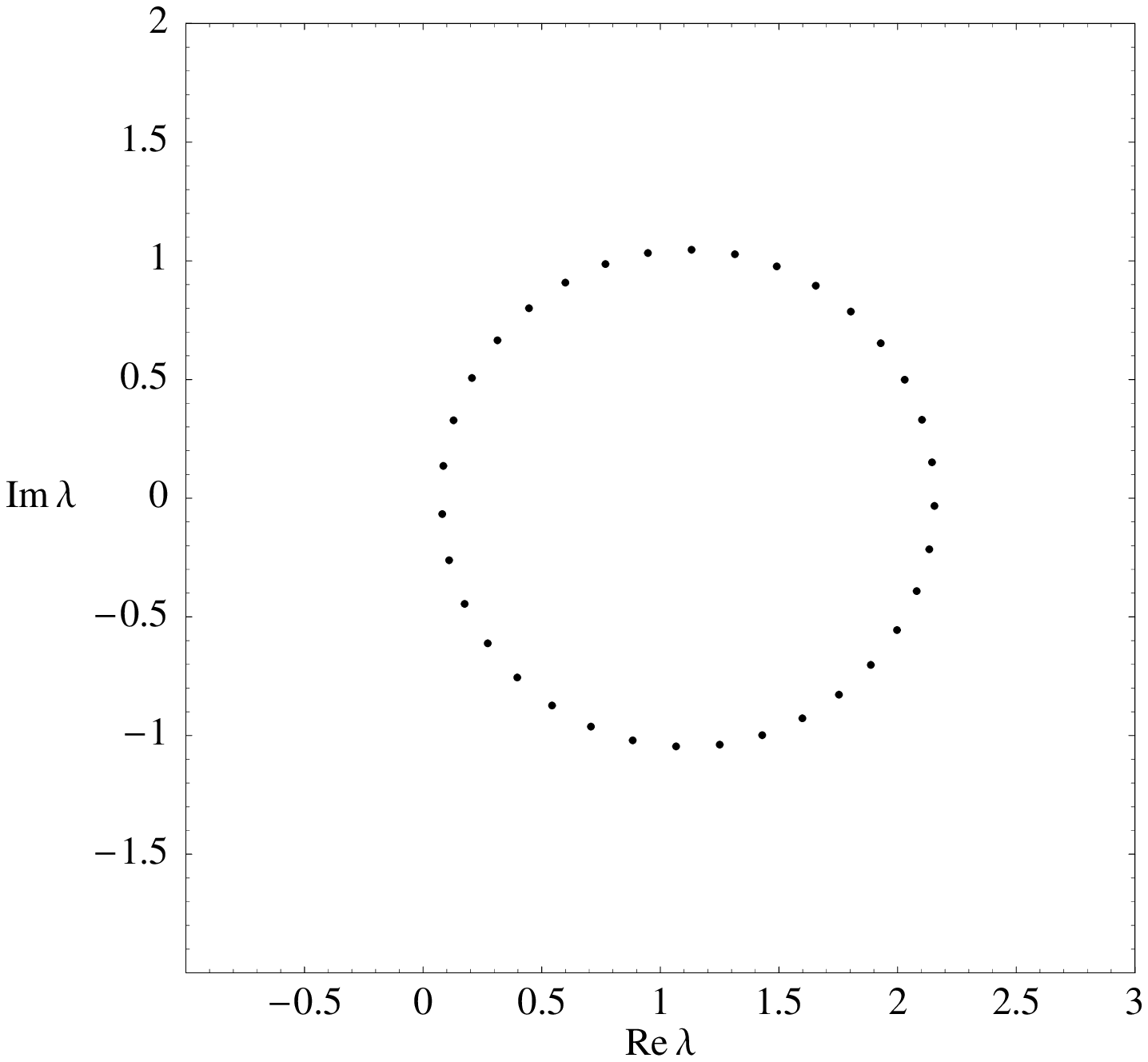,width=7cm}
\end{center}

\noindent {\bf Figure 4:} Eigenvalue distribution of the
Polyakov-Maldacena loop on the complex plane. With $N = 35$ and $N
\frac{2}{3\pi} \eta = 1$.

\end{figure}

\subsection{Solutions for small $\k'$}

This subsection looks at the limit $\k' \to 0$. This is the limit
that makes the most immediate contact with standard field theory
computations, where the strict large $N$ limit is taken prior to
calculating multiply wound loops. Hence $k$ rather than $\k'$ is
kept fixed as $N \to \infty$.

To uncover the behaviour of the solution as $\k' \to 0$ it is
necessary to write
\be
\g = \pi - (3 \pi \k'/2)^{1/3} {\bar \g} \,.
\ee
Then to leading order as $\k' \to 0$ we have that $\bar \g$ must
satisfy the equation
\be\label{eq:smallk}
\frac{d}{dr} \left[f(r) \frac{d{\bar \g}}{dr} \right] = 4 {\bar
  \gamma}^4 \left({\bar \gamma}^3 - 1 \right) \,.
\ee
This is a polynomial nonlinear equation, in contrast to the
general equations of motion we have been considering. The change
of variables ${\bar
\g} = z^{1/3}$ and $dr = f(r) dx$ takes the equation to a form
that narrowly misses having the Painlev\'e property. A very
interesting result comes from computing the action on these
configurations. The resulting action may be written
\be\label{eq:twothirds}
S = - \frac{\sqrt{\lambda} \b r_+ k}{2 \pi} + \frac{1}{N^{2/3}}
\frac{\sqrt{\lambda} \b k^{5/3}}{30 \pi^2} \int_{r_+}^{\infty}
dr \left[15 {\bar \g}^8 - 8 {\bar \g}^5 - 5 f(r)
\left(\frac{d{\bar \g}}{dr}\right)^2 \right]+ \cdots
\,.
\ee
The remaining higher order terms are not reliable in the strict
large $N$ limit without including $1/N$ corrections to the
background and the D5 brane action and so on. The first correction
to the collapsed result appears to be reliable because $N^{-2/3} \gg
N^{-1}$. Numerical
integration of (\ref{eq:smallk}) shows that there are solutions
with finite action which run from infinity to the horizon. One further
has the constant solution ${\bar \gamma} = 1$.
If any of these provide the dual for multiply wound Polyakov loops or
loops in higher representations, then
within the validity of our computation we may write
\be\label{eq:polya}
e^{- S_k} = e^{\eta k} \left[ 1 +
B \frac{k^{5/3}}{N^{2/3}} + \cdots \right] \,,
\ee
where $\eta$ is defined in (\ref{eq:eta}) and $B$ is the integral
given in (\ref{eq:twothirds}). In (\ref{eq:polya}) we have assumed a
$1/r$ falloff which implies that boundary terms don't contribute. In
the case of the constant solution ${\bar \gamma} = 1$ one has
to add the boundary
terms which result in a finite total action, as in the previous
subsection (\ref{eq:constaction}).

The appearance of the non analyticity
$N^{-2/3}$ suggests that the simple large $N$
expansion is breaking down. In fact,
(\ref{eq:polya}) is incredibly tantalising because in the vicinity
of a Gross-Witten transition \cite{Gross:1980he}, the large $N$
expansion does break down \cite{Goldschmidt:1979hq} and
corrections need to be computed using a double scaling limit
\cite{Periwal:1990gf, Periwal:1990qb,Liu:2004vy}. Near the
transition point, corrections are given precisely in terms
of $N^{-2/3}$! Note that the nontrivial $-2/3$ scaling does not arise
in the $\k \to 0$ limit of the spatial Wilson loop
\cite{Drukker:2005kx}.

We should comment on the validity of the solutions. There are
two constraints: the requirement of negligible backreaction on the geometry
and the requirement that worldvolume fields vary slowly
\cite{Callan:1997kz}. Adapted to the present situation, the former of
these requires that $\k' \ll N$ and the latter requires $\k'^{1/3} \gg
1/\lambda^{1/4}$. The first is trivially satisfied while the second is
certainly satisfied for the blown up solutions with finite $\k'$
and can be satisfied as $\k' \to 0$ by taking $\lambda$ sufficiently
large.

\noindent {\bf Note Added:} Soon after this paper appeared on the
arXiv two closely related preprints,
\cite{Yamaguchi:2006tq} and \cite{Gomis:2006sb},
were posted. In the zero temperature supersymmetric spatial
Wilson loop context, these papers have clarified the interpretation of
D5 brane solutions. It has been argued that these solutions compute traces of
Wilson loops in the $k$-th antisymmetric representations of
$SU(N)$. Various interesting questions have been thrown up and we hope
to address these in the near future.

\section{Summary and discussion}

In this paper we have studied the expectation value of Polyakov
loops winding the thermal circle $k$ times in ${\mathcal{N}}=4$
SYM theory at strong coupling. Inspired by the success of
\cite{Drukker:2005kx} in computing multiply wound spatial Wilson
loops, we began by searching for dual probe D3 brane
configurations. We have shown that there are no appropriate
solutions. Instead it appears that the correct dual description of
higher Polyakov loops are probe D5 branes. The required
configurations are, unsuprisingly, similar to those dual to the
baryon vertex \cite{Callan:1998iq}. The blowing up of spheres in
the brane worldvolume modifies the na\"\i ve result $\frac{1}{N}
\langle\Tr U^k\rangle = \frac{1}{N^k} \langle \Tr U\rangle^k$.
In the D3 brane case the blowup is mediated by $\k = k
\sqrt{\lambda}/4N$ while D5 brane blowups are determined by $\k' =
k/N$. Only the latter corrections seem to be present for the
Polyakov loop.

There are various different probe D5 brane solutions of interest.
We concentrated on two classes of solutions. Firstly, we studied
solutions in which an $S^4$ in the D5 brane worldvolume is blown
up at a constant size. The explicit nature of these solutions
allowed us to present a numerical eigenvalue distribution. This
distribution however had $\k'$ rather than $k$ held fixed in the
large $N$ limit and may not be directly comparable to the
eigenvalue distributions computed in field theory.

Secondly, we looked at solutions in the limit of small $\k'$, or
equivalently, $N \to \infty$ with $k$ held fixed. We found that
the leading correction to the na\"\i ve multiply wound string
result is of order $N^{-2/3}$. This is a very curious result as
$N^{-2/3}$ is the scaling one finds in the vicinity of a
Gross-Witten phase transition.

Thus while we are not yet in a position to make hard statements about
the behaviour of the Polyakov loop eigenvalue distribution at strong
coupling, and hence the phase structure of the finite temperature
${\mathcal{N}}=4$ theory, we have identified the required degrees
of freedom on the bulk side of the duality and uncovered various hints
of interesting behaviour.

We are left with many further questions. One immediate open
question is to systematically understand the different possible D5
brane configurations and their field theory interpretations. In
particular, one would like to know exactly which of the
configurations, if any, is dual to the multiply wound loop.

The tantalising appearance of $N^{-2/3}$ corrections suggested a
connection with the double scaling limit near a Gross-Witten phase
transition. A computation that could firm up this statement would
be the calculation of $\langle \Tr U^k \rangle$, and other higher
representation loops, in the double scaling limit. One would be
looking for corrections of the form $k^{5/3}/N^{2/3}$.

It is worth bearing in mind that the Polyakov loop being a
composite operator, its expectation value suffers ultraviolet
divergences. The divergences in Wilson loops have been well understood
(\cite{Dumitru:2003hp} and references therein) and result in
an overall multiplicative constant whose origin is the infinite additive
mass renormalisation of a static test quark in the theory.

Finally, another possibly interesting direction for further investigation
would be to compute the traces of the Maldacena-Polyakov loop in a
partially resummed field theory model. Although this is not a
controlled approximation for non supersymmetric loops, it has
recently been seen to capture important physics away from the
perturbative regime in the 't Hooft coupling
\cite{Klebanov:2006jj}.

\section*{Acknowledgements}

We thank Ofer Aharony, Bartomeu Fiol and Shiraz Minwalla for their
criticisms and comments on earlier versions of this work. We would
also like to thank Gert Aarts, Maciej Dunajski, Roberto Emparan,
Gary Gibbons, Rob Myers, Asad Naqvi, Carlos N\'u\~nez, Nemani
Suryanarayana, Rob Pisarski and Toby Wiseman for helpful comments
during the course of our work. SAH is supported by a research
fellowship from Clare college, Cambridge. SPK is supported by a
PPARC advanced fellowship. SAH would like to thank the KITP for
hospitality during an intermediate stage of this project. This
research was supported in part by the National Science Foundation
under Grant No. PHY99-07949.

\end{document}